\documentclass[12pt, letterpaper, titlepage]{article}
\usepackage{fullpage}
\usepackage[margin=1in, nohead]{geometry}
\usepackage{amsmath}
\usepackage{graphicx}
\usepackage{amsfonts}
\usepackage{amssymb}
\usepackage{subfigure}
\usepackage{textcomp}
\usepackage{titlesec}
\usepackage{enumitem}
\usepackage{nicefrac}
\usepackage{amsmath}
\usepackage{setspace}
\usepackage[mathlines]{lineno}
\usepackage{times}
\usepackage[tablesfirst,notablist]{endfloat}
\usepackage{booktabs, caption, makecell}
\usepackage{multirow}
\usepackage{authblk}
\usepackage{natbib}
\usepackage{soul}

\setcitestyle{authoryear,aysep={}}

\titleformat{\subsection}{\itshape}{\thesubsection}{0em}{}

\makeatletter
    {\if@restonecol\twocolumn \else \newpage \fi
     \if@twoside\else
        \setcounter{page}\@ne
     \fi
    }
\makeatother

\newcommand{\bi}{\boldsymbol{\beta}_{i}}
\newcommand{\y}{y_{i,t}}

\newcommand{\ly}{\text{log}(y_{i,t})}
\newcommand{\bx}{\mathbf{x}_{i,t}'}
\newcommand{\bxnt}{\mathbf{x}_{i,t}}
\newcommand{\stdi}{\alpha_{j(i),t}}
\newcommand{\std}{\alpha_{j,t}}

\newcommand{\err}{\epsilon_{i,t}}

\newcommand{\f}{\mathbf{f}_{j,t}'}
\newcommand{\fnt}{\mathbf{f}_{j,t}}
\newcommand{\stderr}{v_{j,t}}
\newcommand{\procerr}{\mathbf{w}_{t}}

\newcommand{\sigpe}{\sigma_{pe}^{2}}

\newcommand{\intvar}{\tau^2}
\newcommand{\bTh}{\boldsymbol{\theta}}
\newcommand{\bTht}{\boldsymbol{\theta}_{t}}
\newcommand{\bThpt}{\boldsymbol{\theta}_{t-1}}

\title{Variable Effects of Climate on Forest Growth in Relation to Climate Extremes, Disturbance, and Forest Stand Dynamics}
\author{Malcolm S. Itter \thanks{ittermal@msu.edu}\\Department of Forestry, Michigan State University, East Lansing, Michigan 48824 USA\\
Program in Ecology, Evolutionary Biology, and Behavior, Michigan State University, East Lansing, Michigan 48824 USA \and 
Andrew O. Finley \thanks{finleya@msu.edu}\\
Department of Forestry, Michigan State University, East Lansing, Michigan 48824 USA\\
Department of Geography, Michigan State University, East Lansing, Michigan 48824 USA \and 
Anthony W. D'Amato \thanks{awdamato@uvm.edu}\\
Rubenstein School of the Environment and Natural Resources, University of Vermont, Burlington, Vermont 05405 USA \and 
Jane R. Foster \thanks{jrfoster@umn.edu}\\
Department of Forest Resources, University of Minnesota, St. Paul, Minnesota 55108 USA \and 
John B. Bradford \thanks{jbradford@usgs.gov}\\
Southwest Biological Science Center, U.S. Geological Survey, Flagstaff, Arizona 86001 USA}
\date{}

\begin{document}

\maketitle
\addtocounter{page}{1}

\raggedright
\setlength{\parindent}{15pt}

\section{Abstract}
Changes in the frequency, duration, and severity of climate extremes are forecast to occur under global climate change. The impacts of climate extremes on forest productivity and health are complicated by potential interactions with disturbance events and forest stand dynamics. Such interactions may lead to non-linear forest growth responses to climate and disturbance involving thresholds and lag effects. The effects of stand dynamics on forest 
responses to climate and disturbance are particularly important given forest characteristics driven by stand dynamics can be modified through forest management 
with the goal of increasing forest resistance and resilience to climate change. We develop a hierarchical Bayesian state-space model in which climate effects on tree growth are allowed to vary over time and in relation to climate extremes, disturbance events, and stand dynamics. The model is a first step toward integrating stand dynamics into predictions of forest growth responses to climate extremes and disturbance. We demonstrate the insights afforded by the model through its application to a dendrochronology dataset comprising measurements from forest stands of varying composition, structure, and development stage in northeastern Minnesota. Results indicate that average forest growth was most sensitive to water balance variables describing climatic water deficit. Forest growth responses to water deficit over time were partitioned into responses driven by climatic threshold exceedances and interactions with forest tent caterpillar defoliation. Forest growth was both resistant and resilient to climate extremes with the majority of forest growth responses occurring after multiple climatic threshold exceedances across seasons and years. 
Interactions between climate and insect defoliation were observed in a subset of years with forest growth responses to climatic growing conditions in the opposite direction than expected. Forest growth was most sensitive to climate, regardless of driving factor, during periods of high stem density following major regeneration 
events when average inter-tree competition was high. Results suggest that forest growth resistance and resilience to interactions between climate extremes and insect defoliation can be increased through management steps such as thinning to reduce competition during early stages of stand development and small-group selection harvests to maintain forest structures characteristic of older, mature stands.

\subsection*{Keywords}
Bayesian state-space model, climate, disturbance, forest dynamics, forest tent caterpillar, hierarchical models, productivity/biomass, tree rings, dendrochronology

\section{Introduction}
Understanding the effects of climate on forest productivity is integral to predicting the response of forest ecosystems to global climate change. Forest growth responses to changing climatic conditions have important implications for sustainable forest management, a fundamental goal of which is to maintain healthy and productive forests in perpetuity. One projected consequence of  climate change, in addition to global warming trends, is changes in the frequency, severity, and duration of extreme climate or weather events \citep{IPCC2013}. Climate extremes have the potential to profoundly alter the productivity and health of forest ecosystems \citep{Allen2010}.

The increased frequency of droughts combined with warmer temperatures in the Southwestern US, for example, is projected to lead to growth declines in dominant coniferous species and in severe cases large-scale forest mortality \citep{Breshears2005, Williams2010}. Similar growth declines and mortality are projected to occur in mixed oak and pine forests of the Southeastern US and elsewhere around the globe \citep{Klos2009, Berdanier2016}. The impact of droughts on forest health and productivity is compounded with potential interactions between drought and other abiotic and biotic disturbance agents such as wildfire and forest damaging insects \citep{Dale2001}. In particular, the occurrence of droughts may facilitate increased insect populations, or insect damage may exacerbate the effects of drought leading to more severe forest mortality than expected  \citep{McDowell2008, Anderegg2015}. Forest responses to interactions between climate extremes and disturbance are expected to be complex and exhibit non-linear behavior involving lags and threshold effects \citep{Betancourt2004, Williams2010, Macalady2014}.

Forest sensitivity to climate extremes has been shown to vary depending on endogenous forest stand characteristics such as stem density, age, species composition, and developmental or successional stage \citep{Laurent2003, Klos2009, D'Amato2013}. These characteristics are driven by forest stand dynamics \citep[e.g.][]{Oliver1996} and are important for understanding, predicting, and minimizing the effects of climate on forest health and productivity. First, forest stand characteristics are dynamic, altering forest responses to climate extremes and disturbance over time and space. Second, forest stand characteristics can be modified through forest management to increase forest resistance and resilience to changing climatic conditions in the short term and facilitate forest adaptation to climate change in the long term \citep{Millar2007, Puettmann2011}. Effective forest management in the face of global climate change requires understanding and predicting changes in the complex interactions between climate, forest disturbance, and forest stand dynamics \citep{Dale2001}. New analytical approaches capable of dealing with expected non-linear forest responses to climate extremes and disturbance that change over time and space are needed to achieve such understanding and predictive ability \citep{Betancourt2004}.

We develop a hierarchical Bayesian state-space model that allows the effects of climate variables on radial tree growth to vary over time and in relation to climate extremes, forest disturbance, and forest stand characteristics. We demonstrate the additional insight made possible by the model regarding interactive 
effects of climate extremes, disturbance, and stand dynamics on forest growth responses to climate through its application to a dendrochronology dataset from northeastern Minnesota. The dataset includes radial growth measurements from individual trees located in 35 forest stands of varying age, structure, species composition, and developmental stage. The phenomenological model developed herein represents a first step toward a process-based model to 
predict forest growth responses to climate extremes and disturbance as a function of stand characteristics that can be used to guide 
management actions to maintain forest health and productivity under changing climatic conditions.

\section{Modeling Approach}

\subsection*{Background}
Tree rings are valuable observational data for understanding the effects of climate extremes and disturbance on forest productivity. They provide long-term records (centennial to millennial) of radial tree growth that can be used to infer extreme climate/weather and disturbance events \citep{Cook1990}. Growth rings also have a functional relationship to forest basal area and above-ground biomass, common measures used to quantify forest productivity \citep{Babst2014}. Dendrochronology provides a diverse and extensive set of methodologies to infer relationships between inter-annual climate variability and growth rings \citep{Cook1987, Cook1990}. In the current analysis, we seek to develop an inferential approach that allows for estimation of dynamic growth effects of climate and how such effects covary with climate extremes, forest disturbance, and stand dynamics.

Several established dendrochronology methods allow the effects of climate on tree growth to vary over time. A number of studies have developed applications of the Kalman filter to estimate time-varying climate response functions \citep{Visser1986, Visser1988, VanDeusen1989}. Other studies have applied moving correlation analysis to estimate time-dependent correlation coefficients between climate variability and tree growth using a moving window through time \citep{Biondi1997,Biondi2000,Carrer2006}. The Kalman filter approach, in particular, has been applied to identify changes in the effects of climate on tree growth attributable to air pollution \citep{Innes1989}, and to interactions with forest stand dynamics \citep{VanDeusen1987}. Further, the Kalman filter has been shown to capture low-frequency changes in climatic growth effects due to long-term climate shifts and high-frequency changes in growth responses to climate extremes \citep{Visser1988}. Despite its potential, the Kalman filter approach to tree ring analysis is not widely applied in contemporary forest ecology analyses.

Here we offer a modern update of the Kalman filter approach to tree ring analysis in the context of understanding the interactive effects of climate extremes, forest disturbance, and stand dynamics on forest growth. We nest the Kalman filter within a hierarchical Bayesian state-space framework that affords several benefits over past tree ring applications. First, the hierarchical Bayesian approach allows for uncertainty propagation throughout model components. Uncertainty sources include tree ring detrending/standardization, climate growth response functions, and the evolution of climate effects through time. Second, we are able to estimate observation variance (error in composite growth estimates) and process variance (error in temporal evolution of climate effects) coincident with all other model parameters and accommodate non-Gaussian error structures as opposed to conventional Kalman filter approaches that assume error terms are known and Gaussian \citep{Carlin1992}. Finally, the hierarchical Bayesian approach allows for integration of climate extremes, disturbance, and stand dynamic factors to predict changes in climatic effects on forest growth using, potentially non-linear, process equations.

The benefit of hierarchical Bayesian state-space models to understand and predict dynamic ecological processes has been demonstrated in studies of biodiversity \citep{Clark2012}, population dynamics \citep{Calder2003, Clark2004, Parslow2013}, and invasive species \citep{Wikle2003, Hooten2007}. Hierarchical state-space models have also been used to assimilate growth ring and diameter tape measurements of radial tree growth \citep{Clark2007, Schliep2014}. Finally, recent dendrochronology studies have applied hierarchical Bayesian models to improve quantification of uncertainty in tree ring standardization and estimation of growth-climate relationships for climate reconstruction \citep{Schofield2015}.

\begin{figure}
 \begin{center}
 \includegraphics[trim=0in 0.6in 0in 4.4in,clip,width=\textwidth]{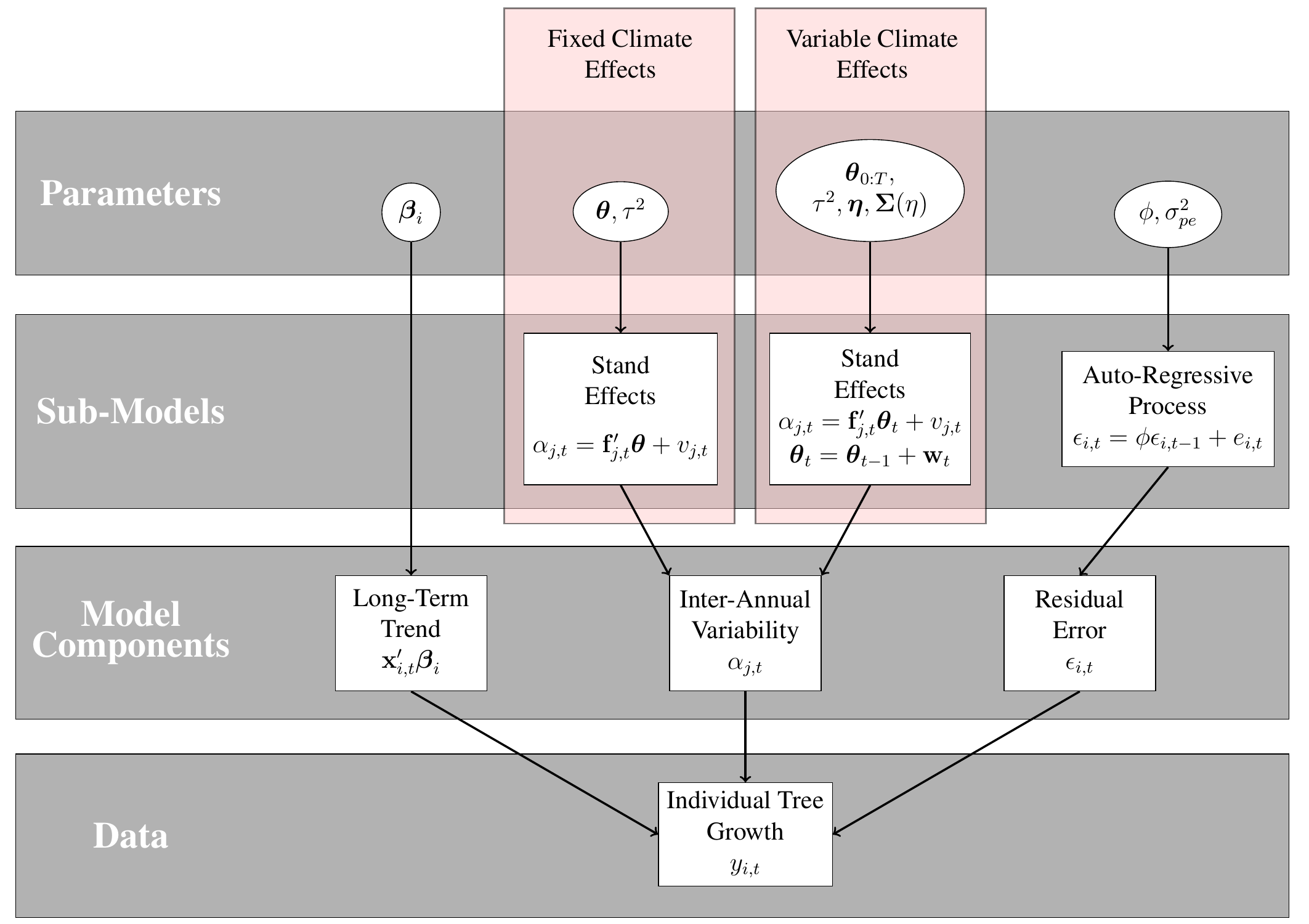}
 \caption[Schematic of fixed climate effects (FCE) and variable climate effects (VCE) tree growth models. \textit{Note:} Fixed and variable climate effects sub-models are applied 
 separately; 0:$T$ subscript indicates all time points from 0 to $T$.]{}
 \label{fig1}
\end{center}
\end{figure}

We apply two models to estimate tree growth as a function of climate. The first model moves the analytical steps involved in a dendrochronology response function analysis, where the objective is to estimate the average effects of a set of climate variables on annual tree growth 
over a fixed period, into an integrated hierarchical Bayesian model allowing for explicit error propagation 
across steps. In particular, application of a Bayesian model-based approach, such as the one developed herein, leads to estimates of climate effects that reflect the uncertainty associated with all model components \citep{Schofield2015}. The second model extends the first allowing climate effects on tree growth to vary annually over the study period using a hierarchical Bayesian state-space approach. Throughout, we refer to the first model as the \emph{fixed climate effects} model, and the second model as the \emph{variable climate effects} model. Sources of uncertainty under the fixed climate effects model include steps to detrend/standardize individual tree growth records, generate composite growth records or chronologies, and random variability in individual and composite growth 
records. The variable climate effects model shares the same sources of uncertainty as the fixed climate effects model with additional uncertainty associated with the evolution of climate effects through time. An overview of the fixed and variable climate effects models 
are included in the following two subsections. Figure \ref{fig1} provides a schematic representation of the two models. 
The statistical details and discussion of Bayesian inference applied to estimate model parameters are provided in 
Appendix A.

\subsection*{Fixed Climate Effects}
We begin by decomposing individual tree growth into component sources of variability to explain the different submodels included in the fixed climate effects (FCE) model. Specifically,
\begin{center}
 \includegraphics[trim = 1in 9.55in 2.5in 1in, clip=true]{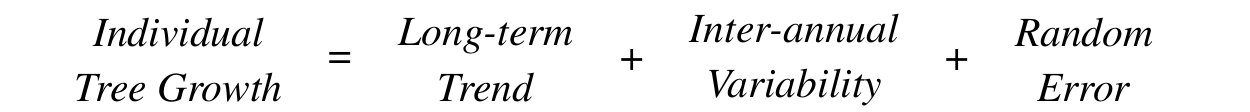}
\end{center}
where the long-term trend captures low-frequency changes in growth due to tree size and age, 
the inter-annual variability captures high-frequency changes in growth potentially driven by climate, 
and the random error term captures residual variability.

\textbf{Long-Term Trend}: A smoothing spline is commonly applied in dendrochronology to model the long-term, 
low-frequency trend in individual growth attributable to tree size and age \citep{Cook1981}. Smoothing splines 
are a highly flexible method to model natural phenomena using a set of polynomial basis functions \citep{Wood2006}. 
We apply a penalized spline regression submodel in the FCE model to capture long-term size and age effects, although 
a variety of smoothing approaches can be used here (see Appendix A for a complete discussion).

\textbf{Inter-Annual Variability}: High-frequency, inter-annual variability in tree growth records is frequently attributed to 
the effects of climate and forest stand dynamics \citep{Cook1987}. We model mean inter-annual growth variability 
with an additive annual stand effect that represents the average deviation of each tree within a stand from the tree's long-term 
growth trend within a given year. Inter-annual variation in the additive stand effect is in turn 
modeled as a function of observed climate. We model the mean variation rather than individual variation because individual trees within 
a forest stand are likely to exhibit differential responses to climate variability and stand dynamics, hence the use of growth chronologies 
in studies of tree growth and climate.

\textbf{Random Error}: The residual error is modeled using an autoregressive process to explicitly account for temporal autocorrelation in 
annual tree growth increments. Specifically, we apply a first order autoregressive (AR1) model for the residual error. We note the residual 
error can also be modeled as a function of the mean growth increment to account for heteroscedasticity. We did not choose to do so here.

\textbf{Log Transformation}: We model annual tree growth increments on the log-scale to ensure positive growth estimates consistent 
with \citet{Clark2007}. Log transforming growth increments results in multiplicative errors equivalent to modeling non-transformed 
growth increments using a negative exponential model (as discussed in \citealt{Schofield2015}). Log transforming growth increments 
also reduces the heteroscedasticity frequently observed in tree ring records.

Combining the submodels and log transformation, we model annual tree growth increments in a hierarchical Bayesian model as follows. 
Let $i$ index individual trees ($i=1,\ldots,n$), $j$ index stands ($j=1,\ldots,k$), and $t$ index years ($t=1,\ldots,T$) where $n$, $k$, and $T$ are 
the total number of trees, stands, and years, respectively. Let $j(i)$ indicate the stand $j$ in which the 
$i$th tree is located (e.g. $j(i) = 3$ indicates the $i$th tree is located in stand 3). Finally, define $y$ to be the observed growth 
increment, such that $\y$ is the observed growth increment for tree $i$ in year $t$. Individual tree growth is modeled as,
\begin{linenomath*}
\begin{equation}
 \underbrace{\ly}_{\substack{\text{log-transformed}\\\text{growth increment}}} = 
 \underbrace{\bx\bi}_{\substack{\text{long-term}\\\text{trend}}} + 
 \underbrace{\stdi}_{\substack{\text{inter-annual}\\\text{variability}}} + 
 \underbrace{\err}_{\substack{\text{random}\\\text{error}}} \label{treemod}
\end{equation}
\end{linenomath*}
where $\bxnt$ includes tree age covariates, $\bi$ is a set of tree-specific regression coefficients, $\stdi$ is the additive 
effect of being located in stand $j$ during year $t$, and $\err$ is the residual error modeled as an AR1 process. We 
assume the error term is normally distributed, $\err \sim \mathcal{N}\left(0,\nicefrac{\sigpe}{(1-\phi^{2})}\right)$, where 
$\sigpe$ is the pure error variance and $\phi$ is the temporal autocorrelation coefficient. Stand effects 
reflecting mean inter-annual growth variability across all trees in a stand are modeled using observed climate as,
\begin{linenomath*}
\begin{equation}
 \std = \f\bTh + \stderr \label{stdmod}
\end{equation}
\end{linenomath*}
where $\fnt$ includes observed, standardized climate covariates, 
$\bTh$ is a set of stand-level regression coefficients, and $\stderr$ is a random error term assumed to be 
independent with respect to time both within and across stands, $\stderr \sim \mathcal{N} \left(0,\intvar\right)$.

\subsection*{Variable Climate Effects}
The variable climate effects (VCE) model extends the FCE model to allow climate regression coefficients ($\bTh$ in equation \ref{stdmod}) 
to vary over time. The climate regression coefficients are treated as state variables in the VCE model and evolve over time such that 
a unique set of climate coefficients is estimated for each year in the study period ($\bTh_{1},\bTh_{2},\ldots,\bTh_{T}$). Annual 
climate coefficient estimates are updated using the Kalman filter and are informed by coefficient values for the previous and subsequent years ($t-1$, $t+1$) and annual 
stand effect estimates (Fig. A1). Annual climate coefficients are estimated using stand effects for a five-year period centered 
on the current year ($t-2$ through $t+2$). The use of a five-year moving window allows for partial temporal pooling of tree growth 
data similar to a moving correlation analysis \citep{Biondi1997}, and provides increased sample size to estimate annual climate effects.

The tree-level model (equation \ref{treemod}) is unchanged in the VCE model. The stand-level model (equation \ref{stdmod}) 
is updated to integrate time-varying climate coefficients.
\begin{linenomath*}
\begin{equation}
 \std = \f\bTht + \stderr \label{stdmodvar}
\end{equation}
\end{linenomath*}
The evolution of climate coefficients over time is modeled using a random walk.
\begin{linenomath*}
\begin{equation}
 \bTht = \bThpt + \procerr \label{ccmod}
\end{equation}
\end{linenomath*}
We assume both the stand effect error and random walk error terms follow normal distributions, 
$\stderr \sim \mathcal{N}(0,\intvar)$, and $\procerr \sim \mathcal{N}(\mathbf{0},\boldsymbol{\Sigma}_{\theta})$. 
Equations (\ref{stdmodvar}) and (\ref{ccmod}) define a state-space or dynamic linear model framework with 
(\ref{stdmodvar}) serving as the observation equation and (\ref{ccmod}) the process or state equation 
\citep{West1997}. Details on the state-space modeling approach and the numerical methods used to 
estimate model parameters including the application of the Kalman filter are provided in Appendix A.

\textbf{Criteria for evaluating variable climate effects}: Time-varying climate coefficients can be difficult 
to interpret. Understanding the factors that contribute to a significant forest growth response to a climate variable
in a given year is especially challenging, particularly when multiple climate variables are considered. We applied 
the following criteria to evaluate time-varying climate coefficients and facilitate interpretation of VCE model 
results. A climate variable was considered to have no effect on mean annual growth at the 
stand scale in a given year if the 95 percent credible interval for the variable included zero. A climate variable was 
considered to have only a weak effect in a given year if its 95 percent credible interval did not include zero, but the climate model 
explained less than 25 percent of the variability in the five years of annual stand effects centered on the current year 
(i.e. annual $r^2 < 0.25$). Mean annual growth at the stand scale was considered sensitive to a climate variable in a given year if the 
95 percent credible interval for the variable did not include zero, and the climate model explained at least 25 percent of the variability in the five years of annual stand effects centered on the current year (in the remainder we use forest growth 
response to describe variable by year combinations for which forest growth sensitivity was observed).

We partitioned years during which there was evidence of growth sensitivity for one or more 
climate variables into four categories. First, we defined climatic thresholds as the upper quantiles of observed mean annual 
climate values across stands (quantile values used varied depending on the climate variable to ensure separation of forest growth responses; Fig. \ref{fig5}). Growth sensitivity to a climate variable in a given year was attributed to a 
threshold exceedance if the threshold for a variable was exceeded within the five-year moving window of annual stand effects 
used to estimate climate coefficients. Growth sensitivity to a climate variable was attributed to a persistent exceedance effect if 
there were continued years of growth sensitivity following the five-year period centered on a climatic threshold exceedance. Growth 
sensitivity was attributed to interactions with the forest tent caterpillar if years of growth sensitivity coincided with caterpillar 
outbreak years for known host species. Finally, 
growth sensitivity to a climate variable in a given year was attributed to unknown sources if it did not 
meet any of the previous criteria.

\section{Data}
\subsection*{Tree Growth Data}
\begin{figure}
 \centering
 \includegraphics[scale=0.5]{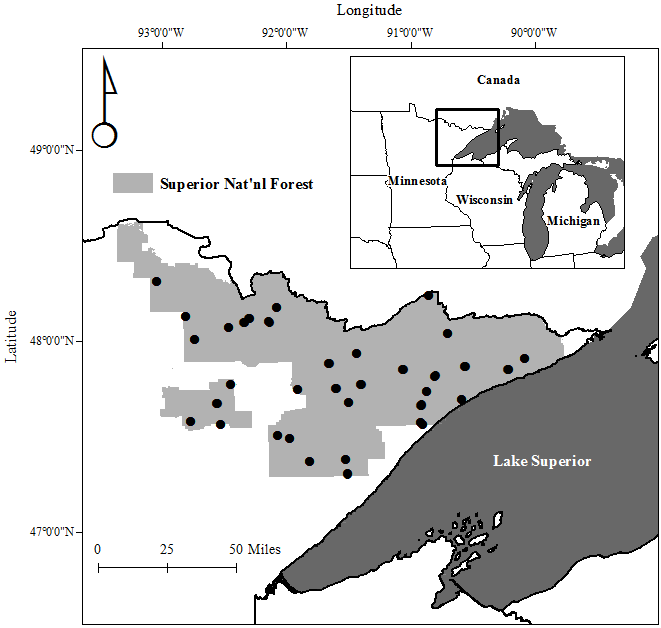}
 \caption[Location of 105 forest plots (35 stands, 3 plots per stand) in relation to Superior National Forest in northeastern 
 Minnesota, USA.]{}
 \label{fig2}
\end{figure}

Tree growth data was collected in 2010 from 35 forest stands in and around Superior National Forest in 
northeastern Minnesota \citep[Fig. \ref{fig2};][]{Foster2014}. Stands were selected to represent 
the predominant forest communities in the broader geographical area based on National Forest Inventory and 
Analysis data from 2004 to 2008 and included a mixture of species compositions, age structures, and development stages. The study region spans the temperate-boreal forest ecotone and sampled forest types reflected this biogeographic setting ranging from common temperate forest types, such as \textit{Acer saccharum}-dominated northern hardwood forests, to boreal forests dominated by \textit{Pinus banksiana}, \textit{Populus tremuloides}, and \textit{Picea spp}. Three replicate 400-m$^2$ circular plots were established within each stand. 
Increment cores were collected at breast height (1.3 m) from all live trees with a diameter at breast height (DBH) larger than 
10 cm. Increment cores were measured using a Velmex measuring stage and crossdated according to standard dendrochronological 
techniques \citep{Holmes1983, Yamaguchi1991}. The most-recent year in which growth data was available for all study trees was 2007.

The DBH and species of sample trees were recorded and tree locations were mapped (relative to plot center coordinates). 
Tree age was estimated by defining the pith as the recruitment year and counting the number of growth rings from recruitment to present. 
Data suitable for climate modeling exist for 2,291 trees representing 15 unique species located across 105 forest plots (Tab. \ref{Tab1}). Sampled forest stands are assumed to 
have established following a stand-replacing disturbance event (e.g. timber harvest or fire); however, age 
was estimated by setting the year of establishment equal to the 25th percentile of the tree recruitment year distribution for each stand to 
account for extended periods of multi-age forest conditions \citep{Foster2014}. 
The start of the study period was set to 1897 to be consistent with the earliest available climate data, although the growth records for a 
subset of trees date back before 1897 (study period = 1897 to 2007). As is the case with nearly all dendrochronology datasets, 
the number of trees and stands observed each year increases with time reflecting trees that established between the 
start and end of the study period.

\begin{table}
\caption{Summary of species sampled.}
\label{Tab1}
\begin{centering}
\begin{tabular}{lccccc}
  \hline\hline
\multirow{2}{*}{Species} & Species & Number of & Number of & First Year & Mean \% Relative \\
& Code & Trees & Plots & Observed & Basal Area\\
  \hline
  \textit{Abies balsamea} & ABBA & 365 & 59 & 1897 & 9 (0, 62) \\ 
  \textit{Acer rubrum} & ACRU & 87 & 30 & 1899 & 3 (0, 23) \\ 
  \textit{Acer saccharum} & ACSA & 175 & 16 & 1899 & 9 (0, 97) \\ 
  \textit{Betula papyrifera} & BEPA & 273 & 50 & 1897 & 13 (0, 70) \\ 
  \textit{Fraxinux nigra} & FRNI & 132 & 9 & 1897 & 7 (0, 97) \\ 
  \textit{Larix laricina} & LALA & 10 & 6 & 1903 & 1 (0, 16) \\ 
  \textit{Picea glauca} & PIGL & 96 & 30 & 1929 & 5 (0, 50) \\ 
  \textit{Picea mariana} & PIMA & 400 & 36 & 1897 & 13 (0, 98) \\ 
  \textit{Pinus banksiana} & PIBA & 383 & 23 & 1919 & 13 (0, 93) \\ 
  \textit{Pinus resinosa} & PIRE & 33 & 9 & 1903 & 3 (0, 47) \\ 
  \textit{Pinus strobus} & PIST & 56 & 15 & 1898 & 5 (0, 77) \\ 
  \textit{Populus grandidentata} & POGR & 23 & 4 & 1927 & 2 (0, 33) \\ 
  \textit{Populus tremuloides} & POTR & 93 & 25 & 1925 & 6 (0, 66) \\ 
  \textit{Quercus rubra} & QURU & 118 & 11 & 1919 & 7 (0, 91) \\ 
  \textit{Thuja occidentalis} & THOC & 47 & 12 & 1897 & 6 (0, 68) \\ 
   \hline
  \multicolumn{6}{l}{\begin{minipage}{6in}
  \textit{Notes:} Number of plots is the number of plots in which each species is found out of a maximum of 105 plots. First year observed is the 
  first year in the study period a growth record exists for a tree of the corresponding species (several trees have records that date back 
  prior to 1897). Mean percent relative basal area is calculated across all study stands based on tree diameters in 2007; the minimum and 
  maximum percent relative basal areas across study stands are provided in parentheses.
  \end{minipage}} \rule{0pt}{3.5\normalbaselineskip}
\end{tabular}
\end{centering}
\end{table}

\subsection*{Forest Tent Caterpillar}
The forest tent caterpillar (\emph{Malacosoma disstria}) is an important native defoliating insect in northeastern Minnesota. There have been several 
forest tent caterpillar (FTC) outbreaks in the region during the study period. Most notably, FTC outbreaks resulted in significant defoliation of 
susceptible trees during the following periods: 1951-1959; 1964-1972; 1989-1995; 2000-2006 \citep{Reinikainen2012}. FTC defoliation in 2001 was 
particularly severe with greater than 7.5 million acres of susceptible hardwood forests in the state suffering defoliation \citep{MNDNR2014}. Study species 
that are known FTC hosts include: \emph{Acer saccharum}, \emph{Betula papyrifera}, \emph{Populus grandidentata}, \emph{P. tremuloides}, and \emph{Quercus rubra}.

\subsection*{Climate Data}
Mean monthly temperature and precipitation estimates were obtained for the study period at a 4-km resolution from 
PRISM \citep{prism}. Climate data were assigned to individual stands by intersecting plot centroids with the PRISM grid and averaging climate observations across the three stand plots if they fell within different grid cells (occurs for 3 out of 35 stands). 
A number of studies have shown that temperature and precipitation are poorly correlated with plant distribution and growth in comparison 
to water balance metrics that translate raw climate observations into variables with direct physiological relevance to plant function 
\citep{Stephenson1998, Dyer2004, Lutz2010}. We derived monthly values of potential evapotranspiration (PET), 
actual evapotranspiration (AET), climatic water deficit (DEF: PET - AET), and mean snow pack using a modified Thornthwaite-type water balance model 
(\citealt{Lutz2010}; see Appendix B for details). We calculated seasonal aggregations for each variable where relevant as 
detailed in Table \ref{Tab2} for a total of 28 climate variables.

\begin{table}
 \caption{Summary of seasonal aggregations of climate variables. Bullets indicate that a seasonal aggregation is 
 calculated for a given variable.}
 \label{Tab2}
 \begin{centering}
  \begin{tabular}{c|c|c|c|c|c}
   \hline\hline
   \multirow{2}{*}{Variable} & Fall & Winter & Spring & Summer & Summer Lag\\
   & (Sep - Nov)$_{t-1}$ & (Dec - Feb)$_{t}$ & (Mar - May)$_{t}$ & (Jun - Aug)$_{t}$ & (Jun - Aug)$_{t-1}$\\ \hline
   Mean Tmin & \textbullet & \textbullet & \textbullet & \textbullet & \textbullet \\ \cline{2-6}
   Mean Tmean & \textbullet & \textbullet & \textbullet & \textbullet & \textbullet \\ \cline{2-6}
   Mean Tmax & \textbullet & \textbullet & \textbullet & \textbullet & \textbullet \\ \cline{2-6}
   Total AET & \textbullet &  & \textbullet & \textbullet & \textbullet \\ \cline{2-6}
   Total PET & \textbullet &  & \textbullet & \textbullet & \textbullet \\ \cline{2-6}
   Total DEF & \textbullet &  & \textbullet & \textbullet & \textbullet \\ \cline{2-6}
   Mean SNOW & & \textbullet &  &  &  \\
   \hline
   \multicolumn{6}{l}{\begin{minipage}{6.25in}
   \textit{Notes:} Tmin, Tmean, Tmax indicate minimum, mean, and maximum temperature, AET and PET indicate actual and potential evapotranspiration, 
   DEF indicates climatic water deficit (PET - AET), SNOW indicates snow pack. Subscripts, $t =$ year of growth, $t-1 =$ year preceding growth.
   \end{minipage}} \rule{0pt}{2.5\normalbaselineskip}
  \end{tabular}
 \end{centering}
\end{table}

\textbf{Bayesian Variable Selection}: We applied the Bayesian Lasso to select a reduced set of climate variables with the greatest effect on 
annual tree growth \citep{Park2008}. While not a formal model-based variable selection technique, the Bayesian Lasso shrinks the coefficient 
values for unimportant variables to zero in a regression model. We chose to apply the Bayesian Lasso given its ability to accommodate collinear 
variables since many of our climate variables were correlated. Additional details on the Bayesian Lasso and its implementation are provided in 
Appendix A. Applying the Bayesian Lasso, the 28 climate variables were pared down to a final set of five climate variables: fall deficit (FAL-DEF), spring deficit (SPR-DEF), summer deficit 
(SUM-DEF), summer deficit in the previous growing season (SUM-DEF-LAG), and mean annual snow pack (SNOW). All model results are 
restricted to this final set of climate variables. Intuitively, we expected water deficit variables to have a negative effect on tree growth. Snow 
pack, reflecting spring soil water recharge given the parametrization of the water balance model (see Appendix B), was expected to have a positive effect on tree growth, though reduced growing season length due to prolonged snow cover could negatively affect growth.

\section{Results}

\subsection*{FCE Model}
\begin{figure}
 \centering
 \includegraphics[scale=0.7]{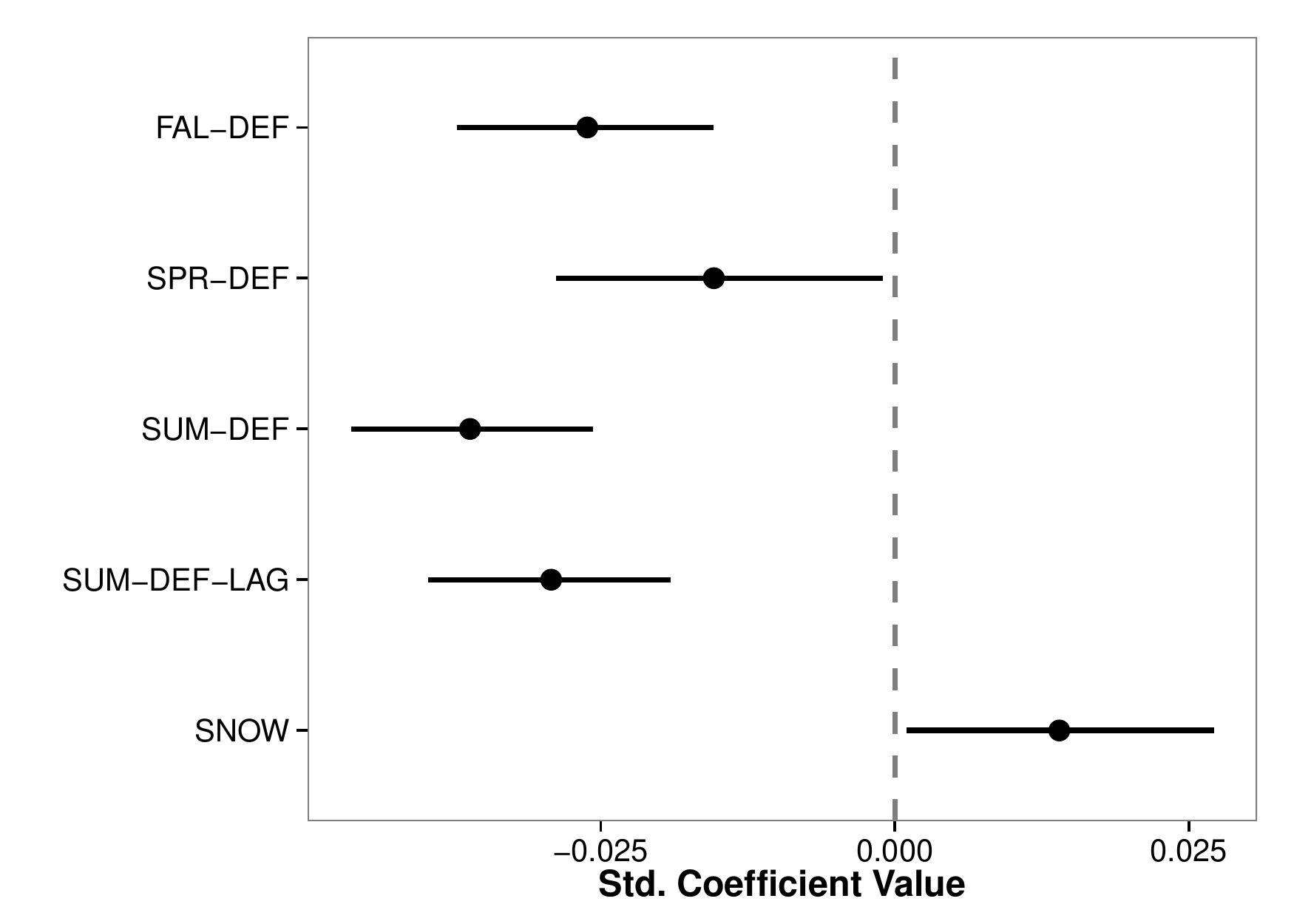}
 \caption[Standardized coefficient values for fixed climate effects (FCE) model. \textit{Note:} Points represent posterior median coefficient estimates; black lines 
 indicate 95 percent credible interval bounds; a dashed grey line at a coefficient value of zero is provided for reference.]{}
 \label{fig3}
\end{figure}

Mean annual stand-level tree growth was sensitive to all five water balance variables in the model as indicated by the 95 percent 
credible intervals not overlapping with zero (Fig. \ref{fig3}). 
Specifically, the four variables representing seasonal climatic water deficit (FAL-DEF, SPR-DEF, SUM-DEF, SUM-DEF-LAG), where larger values indicate greater water deficit, 
were negatively related to mean annual growth at the stand level, while snow pack was positively related to mean annual growth. Climatic deficit 
in the fall, summer, and summer before the year of growth were most related to mean annual growth at the stand level (credible interval bounds are farthest from zero). The climate coefficient estimates in the FCE model represent the average effects 
of the five climate variables over the study period.

Posterior variance estimates for the FCE model are provided in Appendix A (Tab. A1). Notably, the first-order autocorrelation 
coefficient was roughly 0.37 indicating that tree ring records were moderately autocorrelated even after detrending. The 
individual-tree variance was approximately 0.29, while the inter-annual variance (capturing stand-level variability) was 
approximately 0.05, both on the log scale. The individual tree growth variance was roughly six times the stand-level 
growth variance on the log scale.

\subsection*{VCE Model}
\begin{figure}
 \centering
 \includegraphics[width=\textwidth]{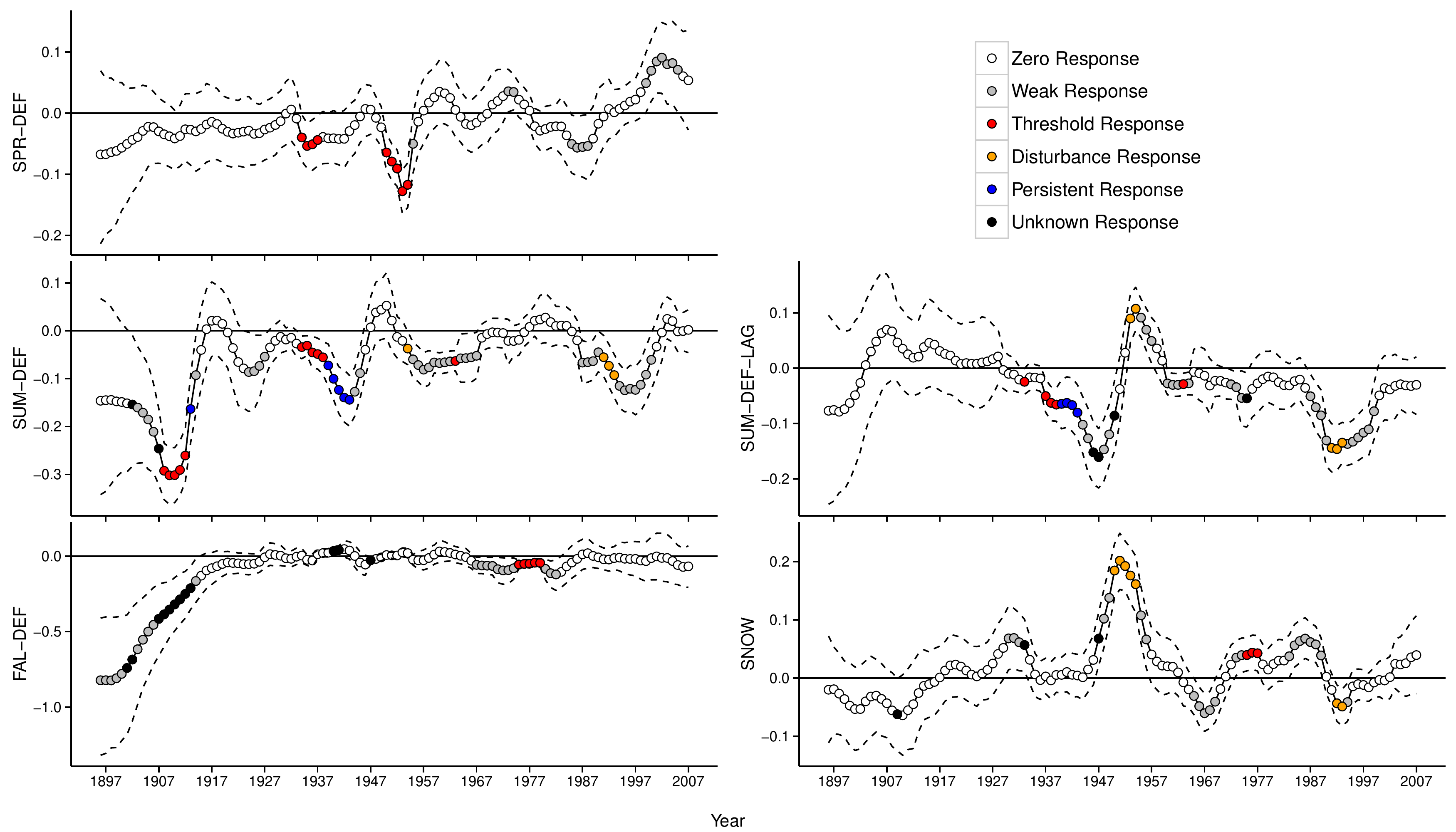}
 \caption[Evolution of climate coefficient values for each climate variable over the study period (1897-2007). Solid black line and 
 points indicate posterior mean coefficient values. Dashed lines delineate posterior 95 percent credible intervals. Points 
 are colored to indicate different response categories. Zero Response: credible interval includes zero; Weak Response: 
 credible interval does not contain zero, but annual $r^2 < 0.25$; Threshold Response: strong response to climate (credible interval 
 does not contain zero and annual $r^2 \geqslant 0.25$) within two years of threshold exceedance; Persistent Response: strong response to 
 climate in years immediately following a threshold exceedance; Disturbance Response: strong response to climate in years 
 where forest tent caterpillar is present in study stands; Unknown Response: strong response to climate not attributable to 
 threshold exceedance, persistent response, or disturbance.]{}
 \label{fig4}
\end{figure}

Estimates of annual effects for each of the five water balance variables are obtained applying the VCE model. The evolution of each variable 
over the study period is presented in Figure \ref{fig4}. While the FCE model demonstrated that mean annual growth at the stand level was 
sensitive to all five water balance variables, there is evidence under the VCE model that the sensitivity of mean annual growth to each 
variable changed in strength and, in some cases, direction over the study period following the sensitivity criteria described in Modeling Approach section.  Table \ref{Tab3} summarizes the results shown in Figure \ref{fig4} partitioning sensitive years 
for each climate variable into each response category. The most common source of growth sensitivity to any climate variable was a 
threshold exceedance, either during the exceedance year ($\sim 41$ percent of all sensitive growth years) or 
in the years following an exceedance ($\sim 13$ percent of all sensitive growth years). Growth sensitivity to one or more climate 
variables coincided with forest tent caterpillar defoliation in 20 percent of all sensitive growth years. There is evidence that trees driving large stand-level growth decreases relative to mean annual 
growth in all study stands (indicative of defoliation) during periods of correspondence between regional forest tent caterpillar defoliation and growth sensitivity to climate were forest tent caterpillar hosts (Fig. \ref{fig6}). Finally, growth sensitivity to one or more climate variables was due to unknown sources (i.e. could not be attributed to a climatic threshold exceedance or forest tent caterpillar defoliation) for roughly 26 percent of all sensitive growth years.

\begin{table}[!ht]
 \caption{Summary of tree growth sensitivity to climate variables (reference for Figure \ref{fig4}).}
 \label{Tab3}
 \centering
  \begin{tabular}{ccccc}
   \hline\hline
   \multirow{2}{*}{Variable} & Threshold & Persistent Response & \multirow{2}{*}{Disturbance} & \multirow{2}{*}{Other} \\
   & Exceedance & Threshold Exceedance & &\\ \hline
   \multirow{2}{*}{SPR-DEF} & 1934-1937 &  \multirow{2}{*}{NA} & \multirow{2}{*}{NA} & \multirow{2}{*}{NA}\\
   & 1950-1954 & & &\\ \hline
   \multirow{3}{*}{SUM-DEF} & 1908-1912 & 1913 & 1954 & 1902\\
   & 1934-1938 & 1939-1943 & 1991-1993 & 1907\\
   & 1963 & & &\\ \hline
   \multirow{3}{*}{SUM-DEF-LAG} & 1933 & 1940-1943 & 1953-1954 & 1946-1947\\
   & 1937-1939 & & 1991-1993 & 1950, 1975\\
   & 1963 & & &\\ \hline
   \multirow{4}{*}{FAL-DEF} & \multirow{4}{*}{1975-1979} & \multirow{4}{*}{NA} & \multirow{4}{*}{NA} & 1901-1902\\
   & & & & 1908-1913\\
   & & & & 1940-1941\\
   & & & & 1947\\ \hline
   \multirow{2}{*}{SNOW} & \multirow{2}{*}{1975-1977} & \multirow{2}{*}{NA} & 1950-1954 & 1909, 1933\\
   & & & 1992-1993 & 1947\\ \hline
   Percent of & \multirow{2}{*}{41.25} & \multirow{2}{*}{12.5} & \multirow{2}{*}{20} & \multirow{2}{*}{26.25}\\
   Growth Responses & & & &\\
   \hline
  \end{tabular}
\end{table}

\begin{figure}
 \centering
 \frame{\includegraphics[width=\textwidth]{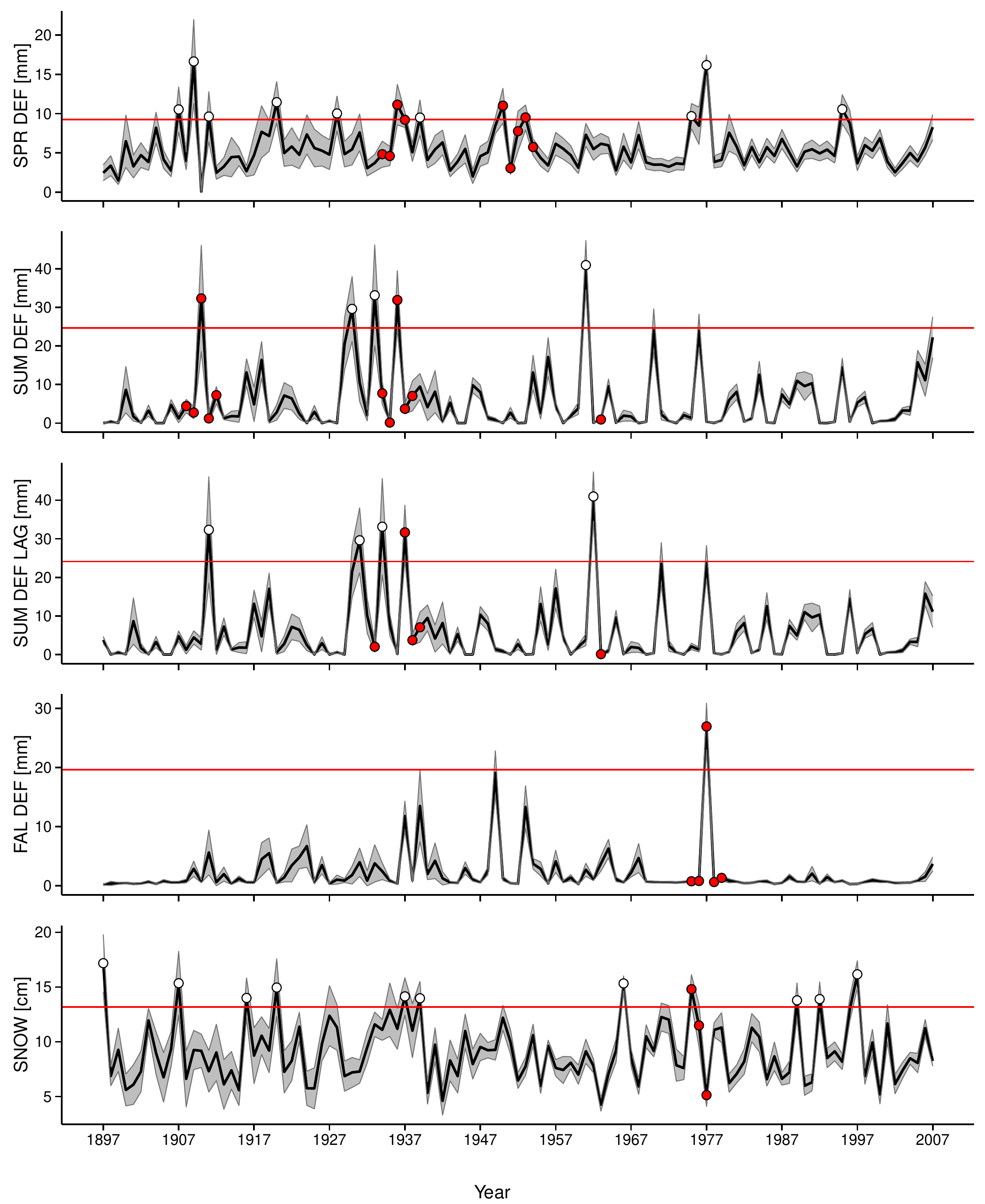}}
 \caption[Observed climate variable values over the study period (1897-2007). Black lines indicate mean climate variable values across 
 study stands along with uncertainty levels equal to two times the standard error (grey shading). The horizontal red line indicates the 
 estimated climate threshold for each variable (thresholds correspond to the following quantiles, 
 0.95: summer deficit, lagged summer deficit; 0.98: fall deficit; 0.85: spring deficit, snow). White filled points indicate threshold 
 exceedances with no growth response. Red filled points indicate years with strong climate growth responses 
 within two years of a climatic threshold exceedance.]{}
 \label{fig5}
\end{figure}

\begin{figure}
 \centering
 \includegraphics[scale=0.85]{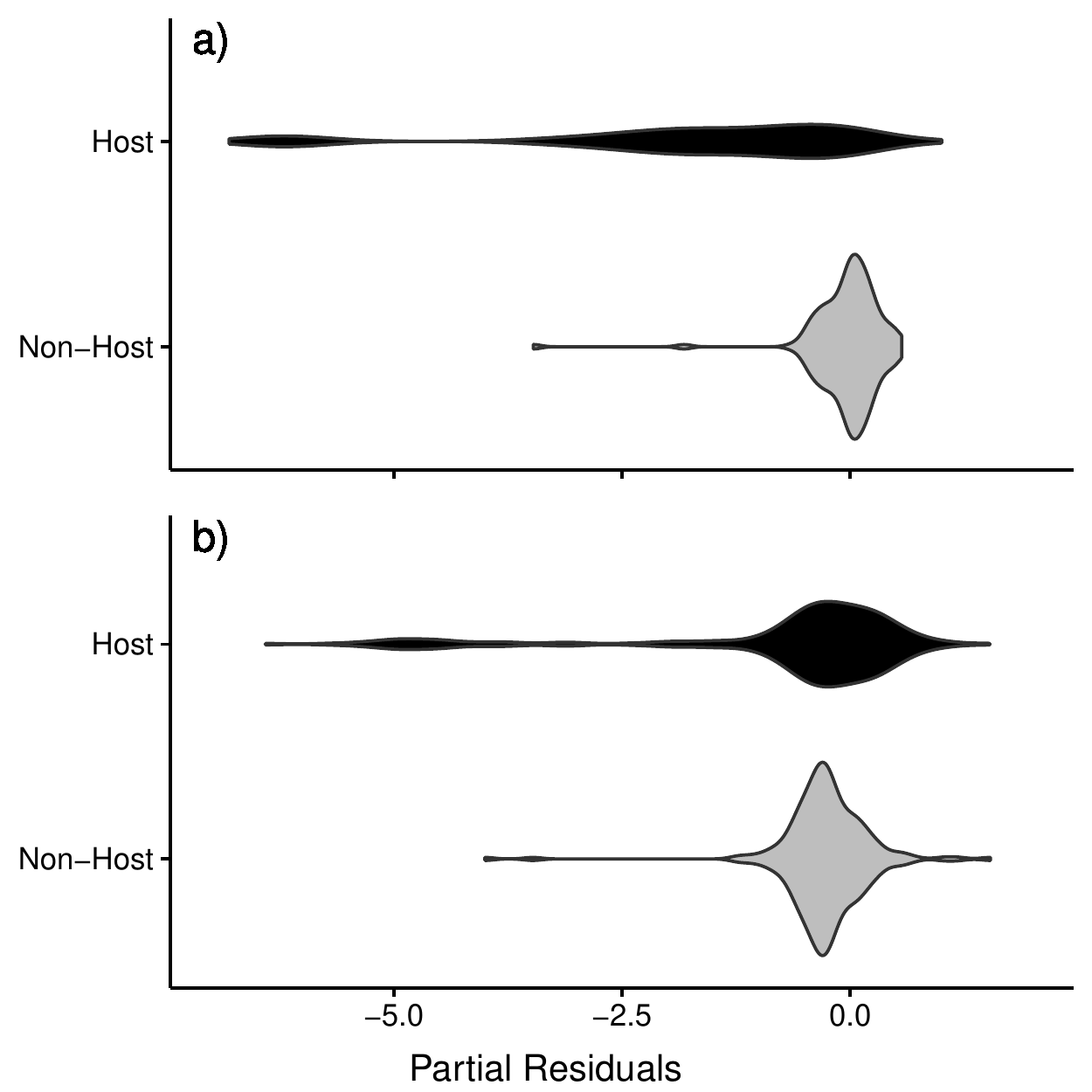}
 \caption[Violin plots of partial residuals (observed log annual growth increment minus spline-based estimate) for forest tent caterpillar (FTC) host and non-host individuals 
 located in stands in the 5th percentile for growth in years a) 1950-1954; b) 1991-1993. Stands in the lowest 5th percentile for growth are 
 considered likely to have been affected by FTC. \textit{Note:} Black shading indicates FTC host trees; grey shading indicates non-FTC host trees.]{}
 \label{fig6}
\end{figure}

Posterior variance estimates for the VCE model are provided in Appendix A (Tab. A1). The variance estimates were consistent 
with the FCE model, except inter-annual stand-level variance, which was slightly smaller under the VCE model (0.042 vs 0.05). 
As in the FCE model, individual-tree growth variance was roughly six times the stand-level growth variance on the log scale. The 
large tree-level variance relative to stand-level variance in both the FCE and VCE models is consistent with previous analyses 
using the same dataset \citep{Foster2016}.

\section{Discussion}

Climate change and associated extreme drought events are expected to fundamentally alter the structure and functioning of forest ecosystems across wide portions of the globe \citep{Clark2016}. The localized impacts of these events on 
forest processes, such as productivity, are likely to vary as a function of tree- and forest stand-level characteristics including species, size, age, and density leading to differential effects across a given landscape and 
over time. Most approaches to modeling climate effects on forest growth have focused on ``average species'' responses 
limiting our understanding of how differences in forest conditions may affect the severity of climate impacts. This 
study presents a modeling framework that underscores the importance of forest dynamics in 
predicting forest growth responses to climate extremes and disturbance, and highlights the potential for management regimes focused on 
manipulating stand structure and density to increase resistance and resilience to future climate change. The current analysis 
focuses on the interactive effects of climate extremes, disturbance, and stand dynamics on forest growth in relatively mesic 
forest sites in northeastern Minnesota. We note the modeling approach developed herein may prove even more useful for 
understanding forest growth responses to drought and disturbance in drier ecosystems such as in the Southwestern US. We begin 
this section with a discussion of the FCE and VCE model results. We then discuss the broader implications of the VCE 
model results to advance understanding of forest growth responses to climate and potential forest management applications.

\subsection*{Fixed Climate Effects}
The Bayesian Lasso allows us to identify the most important climate variables from a large variable set with no \emph{a priori} decisions 
about which variables to consider and without adjustments for collinearity \citep{Hooten2015}. The value of such a tool 
in analyses of the effects of climate on ecological processes is great and has not been previously employed in tree ring analyses. 
We find that water balance variables (climatic water deficit and snow pack) have the largest impact on 
inter-annual tree growth in northeastern Minnesota. Results indicate that tree growth is sensitive to all five water balance 
variables selected by the Bayesian Lasso over the study period with climatic water deficit exhibiting negative growth effects 
regardless of season, and snow pack (a measure of spring soil water recharge) exhibiting positive growth effects. Water 
availability is particularly important in the study region where summers can be quite dry and soils are generally shallow and formed largely from glacial till with poor 
water retention. The climatic water deficit and snow pack variables reflect the interaction 
between temperature, precipitation, and soil water holding capacity (Appendix B). The FCE model results indicate 
that tree growth in the study region is more sensitive to these interactions than to raw temperature and precipitation values 
underscoring the importance of translating climate values into physiologically-relevant variables in studies of tree growth 
as noted in previous studies \citep{Stephenson1998}.

\subsection*{Variable Climate Effects}
The VCE model results suggest that tree growth is sensitive to water balance in punctuated intervals 
of one to several years. We partition these intervals into four categories of potential drivers of growth sensitivity: 
1) exceedance of a climatic threshold indicating a climate extreme; 2) persistent response to a climatic threshold 
exceedance; 3) interaction with disturbance (forest tent caterpillar defoliation); and, 4) unknown sources (Fig. \ref{fig4}, 
Tab. \ref{Tab3}).

\textbf{Climatic Threshold Exceedance}: 
The most frequent driver of tree growth sensitivity is exceedance of a climatic threshold 
representing roughly 41 percent of all observed growth responses. Tree growth responses to water balance threshold 
exceedances are observed in several ways. There are strong responses to singular exceedances, e.g. the observed growth response to summer deficit 
in 1910, a pronounced drought year in the region \citep{Clark1989}, and the response to large fall deficit in 
1977 (Fig. \ref{fig4}). The large fall deficit observed in 1977 coincides with a low mean annual snow pack in 
the same year suggesting forest stands that were water stressed going into the winter received little snowfall to recharge 
depleted soil water in the spring exacerbating the effects of fall deficit. There are also responses to an exceedance 
that closely follow several exceedances in a short period for which no growth response is observed, e.g. the 
observed response to summer deficit in 1936 following two exceedances with no observed response in 1930 and 1933. Considering 
exceedances of all five water balance variables together, we observe several instances of responses to a threshold exceedance 
when exceedances of multiple variables occur coincidentally, e.g. the negative response to fall deficit in 1977 and the positive 
response to a large snow pack in 1975 both coincide with large spring deficits in the same year. The latter example is the only year in which a threshold response 
to snow pack is observed (out of 11 exceedances) suggesting that a positive response to spring water recharge is triggered 
by large spring deficits. There are, however, several years with threshold exceedances of both spring deficit and snow pack 
for which no response to snow pack is observed (1907, 1920, 1939).

In addition to growth responses to a climate variable in the year of a climatic threshold exceedance, we 
observe persistent responses where tree growth is sensitive to a water balance variable for several years following 
a threshold exceedance (12.5 percent of all growth responses; Tab. \ref{Tab3}). We observe persistent responses 
to two large summer deficits (1910, 1936) and to lagged summer deficit in 1937. The persistent response to 
summer deficit in 1936 is noteworthy because the negative effect of water deficit on tree 
growth is largest several years after the exceedance (i.e. persistent response $>$ threshold response). This result 
is consistent with previous dendrochronology analyses in the study region indicating strong radial growth responses 
to lagged moisture availability metrics likely due to depletion of stored, non-structural carbohydrates in years following 
moisture deficits during the growing season \citep{Kipfmueller2010, D'Amato2013}.

Despite evidence that climatic thresholds exist, we cannot establish a definitive threshold for each water 
balance variable above which there is a high probability of a growth response. Rather, we observe a number 
of threshold exceedances for each variable (except fall deficit) with no forest growth response (Fig. \ref{fig5}). 
This may be due to the study period (111 years) being too short to observe sufficient variability in water balance variables 
to establish definitive thresholds. It also suggests that the stands in the study area are relatively resistant to the effects 
of isolated climate extremes. As noted above, responses to threshold exceedances often coincide with multiple exceedances in a 
short period or exceedances of multiple variables. The persistent responses to threshold exceedances are examples of low forest 
growth resilience to climate extremes. We observe persistent responses to only three threshold exceedances suggesting that 
study stands are also relatively resilient to climate extremes.

\textbf{Forest Disturbance}: 
Forest tent caterpillar defoliation appears to affect forest growth responses to climate 
for susceptible host species in the study stands (20 percent of all growth responses coincide with forest tent caterpillar defoliation). Note, in years when the forest tent 
caterpillar is present, stands experiencing defoliation should exhibit more negative stand effects than undisturbed stands. Thus, 
if the growth response to a climate variable is in the expected direction, it implies that the effects of the forest tent 
caterpillar and the climate variable are additive or at least in the same direction. Specifically, stands affected by forest tent caterpillar defoliation 
experience, on average, the poorest climatic growing conditions (e.g. largest climatic water deficits). Conversely, if the growth 
response to a climate variable is in the opposite direction than expected, it implies a strong interaction between forest 
tent caterpillar defoliation and the climate variable or additional unknown factors such that stands affected by defoliation 
experience, on average, better climatic growing conditions than undisturbed stands. In most cases, the response to a climate 
variable in a year with forest tent caterpillar defoliation is in the expected direction implying consistent scaling effects (e.g. 
summer deficit in 1954 and from 1991 to 1993; lagged summer deficit from 1991 to 1993). There are, however, two instances 
where we observe growth responses to a climate variable in the opposite direction than expected: a positive response to lagged 
summer deficit from 1953 to 1954, and a negative response to snow pack from 1992 to 1993. These are the only growth responses to climate variables in 
the opposite direction than expected for the entire study period.

As noted in \citet{Anderegg2015} there can be complex interactive effects of drought stress and insect defoliation stress on forest 
productivity. The positive response to lagged summer deficit in 1953 and 1954 may be due to the presence of drought-weakened trees which were 
subsequently killed or further weakened by caterpillar defoliation creating improved growing conditions for study trees by 
reducing competition levels. The snow pack variable is indicative of spring soil water recharge; large snow pack values, however, may 
cause shortened growing seasons by delaying leaf flush. The negative response to snow pack in 1992 and 1993, therefore, may 
be due to delayed leaf flush reducing carbon assimilation prior to forest tent caterpillar defoliation leading to poor growth years.

\begin{figure}
 \centering
 \includegraphics[scale=0.75]{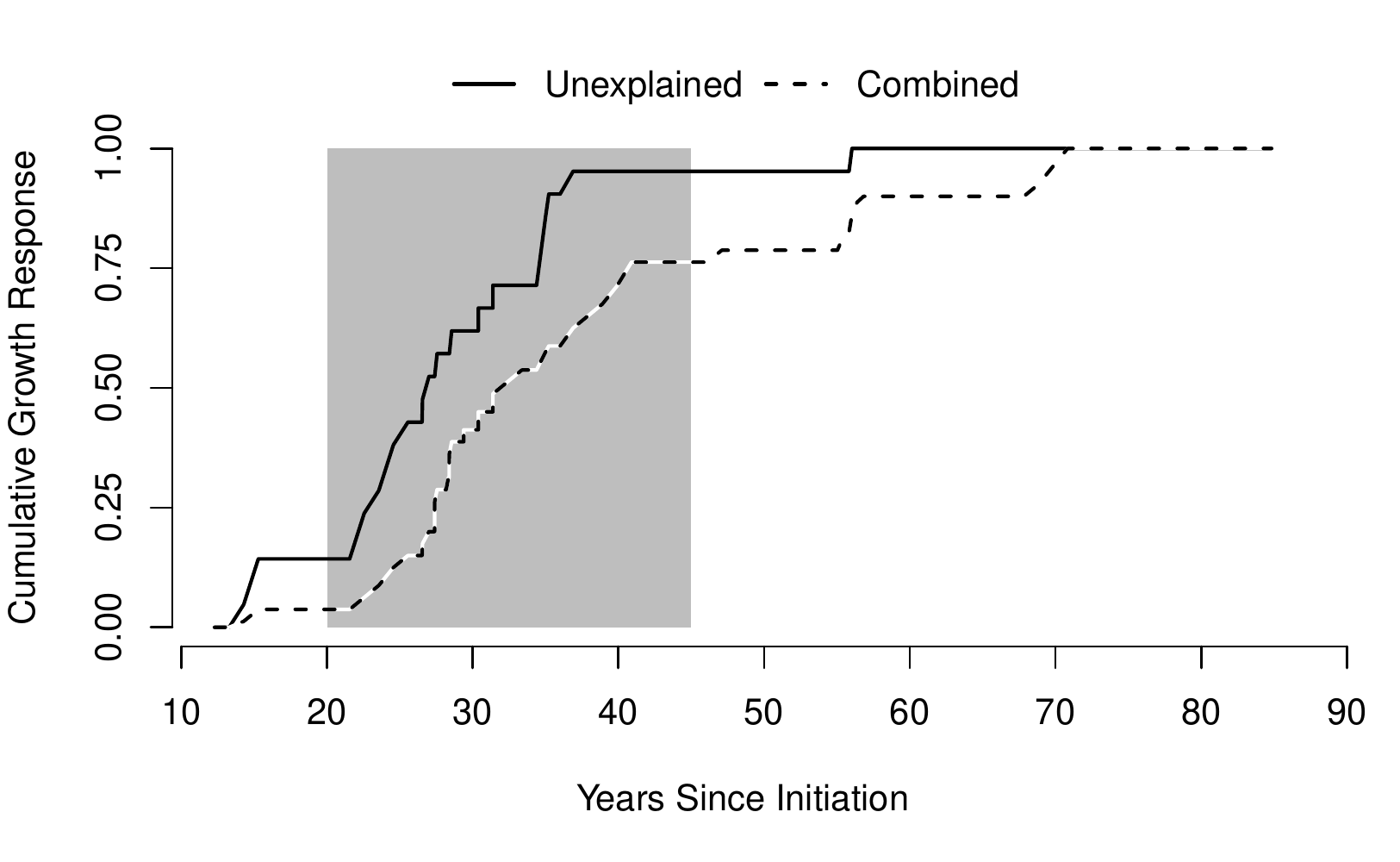}
 \caption[Cumulative distribution of unexplained and combined forest growth responses to water balance 
 variables under the variable climate effects (VCE) model as a function of years since initiation (initiation marks stand establishment in single-cohort stands and 
a large recruitment event in multi-cohort stands). Grey shading highlights stands 20 to 45 years 
 following initiation when understory stem density and inter-tree competition are high.]{}
 \label{fig7}
\end{figure}

\textbf{Stand Dynamics}: 
A number of forest growth responses to one of the five water balance variables cannot be connected to a threshold 
exceedance or forest tent caterpillar defoliation (26.25 percent of all growth responses; Tab. \ref{Tab3}). 
The majority of unexplained forest growth responses occur early in the study period (prior to 1950) when most study 
stands were relatively young. Figure \ref{fig7} depicts the cumulative distribution of unexplained and combined forest 
growth responses as a function of time since stand initiation (note, initiation marks stand establishment in single-cohort stands and 
a large recruitment event in multi-cohort stands). 95 percent of all unexplained forest growth responses occur within 36 years 
of stand initiation. Further, there is a pronounced increase in the number of both unexplained and combined 
forest growth responses between 20 and 45 years since initiation. In single-cohort stands dominated by fast growing, shade-intolerant 
species such as \textit{Populus tremuloides}, \textit{Pinus banksiana}, and \textit{Pinus resinosa}, 20 to 45 years post initiation 
roughly corresponds to the stem exclusion phase of stand development defined by high levels of inter-tree competition and 
density-dependent mortality. Multi-cohort stands, despite not going through the stem exclusion phase, are likely to have high 
stem densities and undergo a high degree of self-thinning in the understory 20 to 45 years following a large regeneration event. 
This suggests the growth of study stands is more sensitive to climate during periods of high stem density following regeneration events when 
individual trees experience higher levels of competition, on average, than in older, mature stands.

This result appears to oppose previous findings indicating growth in older trees is more sensitive to climate than growth in 
younger trees \citep{Carrer2004, Martinez-Vilalta2012}. Although, other studies demonstrating changes in tree growth sensitivity to climate 
due to age have found it is younger tree growth that is more sensitive to climate \citep{Rozas2009}. Previous studies assessing 
the effects of tree age on growth sensitivity to climate utilize growth records from isolated trees in select age classes that minimize 
non-climatic signals. In the current analysis, we use censused tree growth records from stands of varying age, composition, density, 
and structure that are likely to show the effects of past competition and understory growth suppression and release. As such, changes in 
estimates of climate effects on growth over time under the VCE model reflect the evolution of dynamic stand characteristics such as 
age, density, and structure, rather than differences in individual tree ages. 

\textbf{Management Applications}: 
A number of previous studies have demonstrated that forest growth resistance and resilience to drought is sensitive 
to inter-tree competition levels as measured through stand-level basal area. Specifically, there is evidence that 
reducing basal area via thinning increases forest growth resistance and resilience to drought (\citealp{Aussenac1988, 
Laurent2003, Klos2009, Martinez-Vilalta2012, D'Amato2013, Sohn2016}; but see \citealp{Floyd2009} for counter example). 
The benefit of thinning, however, may last only a few years and, in some cases, can cause 
stands to become more sensitive to drought as they mature given an increased presence of large trees with high water demand 
due to large leaf area to sapwood area ratios \citep{McDowell2006, D'Amato2013}. The VCE model results indicate 
that forest growth is most sensitive to climatic water deficit following large regeneration events when forests are 
characterized by high stem densities and high levels of inter-tree competition, particularly in the understory (Fig. \ref{fig7}). 
The current results combined with previous studies suggest that managers may be able to increase forest resistance and 
resilience to climatic water deficit by thinning stands during periods of peak density and inter-tree competition (i.e. stem exclusion 
phase of development in even-aged stands). This period corresponds to the stage at which thinning treatments are traditionally 
applied to increase resource levels for residual trees and mimic density-dependent mortality. Thinning from below 
(removing only trees in intermediate or suppressed canopy positions) may limit the formation of large canopy crowns with high 
leaf area to sapwood area ratios reducing sensitivity to climatic water deficit as stands mature; further, thinning from below 
may minimize levels of evaporative demand at the forest floor due to the high levels of canopy cover it maintains relative 
to other thinning approaches. In multi-cohort stands where intermediate thinning treatments may not be applicable, 
forest managers may be able to increase forest growth resistance and resilience to water deficit by minimizing forest gap sizes 
through individual tree or small group selection harvests that limit the amount of forest in the stem exclusion phase of development 
at a given time and increase the range of tree sizes and spatial diversity present in a given stand. Moreover, the pronounced 
influence of forest tent caterpillar outbreaks and their interaction with climate on productivity underscores the importance 
of maintaining mixed-species stands with a diversity of host and non-host species to minimize the impact of insect defoliation 
on future productivity.

The modeling approach developed in the current analysis relies on the dynamic nature of forests to understand the interactive 
effects of climate extremes and disturbance on forest growth in relation to stand characteristics that may be modified 
through forest management. Future versions of the model will explicitly incorporate stand dynamic factors 
(e.g. development stage, density) and disturbance intensity to predict forest growth responses to interactions 
between climate extremes, disturbance, and stand dynamics and identify management steps to maintain forest 
health and productivity under changing climatic conditions.

\section{Acknowledgments}
The authors thank Christopher Wikle, Giovanni Petris, and Neil Pederson for their helpful comments and feedback during the drafting of this paper. 
This work was supported by National Science Foundation Grants DMS-1513481, EF-1137309, EF-1241874 and EF-1253225 and the Department of Interior Northeast Climate 
Science Center.

\bibliography{TreeGrowthEcoAppAug2016}

\begin{thebibliography}{}

\bibitem[Albers et~al., 2014]{MNDNR2014}
Albers, J., Albers, M., and Cervenka, V. (2014).
\newblock {Forest tent caterpillar fact sheet}.
\newblock {\em Minnesota Department of Natural Resources}.

\bibitem[Allen et~al., 2010]{Allen2010}
Allen, C.~D., Macalady, A.~K., Chenchouni, H., Bachelet, D., Mcdowell, N.,
  Vennetier, M., Kitzberger, T., Rigling, A., Breshears, D.~D., Hogg, E. H.~T.,
  Gonzalez, P., Fensham, R., Zhang, Z., Castro, J., Demidova, N., Lim, J.-h.,
  Allard, G., Running, S.~W., Semerci, A., and Cobb, N. (2010).
\newblock {A global overview of drought and heat-induced tree mortality reveals
  emerging climate change risks for forests}.
\newblock {\em Forest Ecology and Management}, 259:660--684.

\bibitem[Anderegg et~al., 2015]{Anderegg2015}
Anderegg, W. R.~L., Hicke, J.~A., Fisher, R.~A., Allen, C.~D., Aukema, J.,
  Bentz, B., Hood, S., Lichstein, J.~W., Macalady, K., Mcdowell, N., Pan, Y.,
  Raffa, K., Sala, A., Shaw, D., Stephenson, N.~L., Tague, C., and Zeppel, M.
  (2015).
\newblock {Tree mortality from drought, insects, and their interactions in a
  changing climate}.
\newblock {\em New Phytologist}, 208:674--683.

\bibitem[Aussenac and Granier, 1988]{Aussenac1988}
Aussenac, G. and Granier, A. (1988).
\newblock {Effects of thinning on water stress and growth in Douglas-fir}.
\newblock {\em Canadian Journal of Forest Research}, 18(1):100--105.

\bibitem[Babst et~al., 2014]{Babst2014}
Babst, F., Alexander, M.~R., Szejner, P., Bouriaud, O., Klesse, S., Roden, J.,
  Ciais, P., Poulter, B., Frank, D., Moore, D. J.~P., and Trouet, V. (2014).
\newblock {A tree-ring perspective on the terrestrial carbon cycle}.
\newblock {\em Oecologia}, 176(2):307--322.

\bibitem[Berdanier and Clark, 2016]{Berdanier2016}
Berdanier, A.~B. and Clark, J.~S. (2016).
\newblock {Multiyear drought- induced morbidity preceding tree death in
  southeastern U.S. forests}.
\newblock {\em Ecological Applications}, 26(1):17--23.

\bibitem[Betancourt et~al., 2004]{Betancourt2004}
Betancourt, J., D, B., and P, M. (2004).
\newblock Ecological impacts of climate change.
\newblock In {\em Report from a NEON Science Workshop. August 24-25, 2004,
  Tuscon, AZ.} American Institute of Biological Sciences, Washingtion, D.C.,
  USA.

\bibitem[Biondi, 1997]{Biondi1997}
Biondi, F. (1997).
\newblock {Evolutionary and moving response functions in dendroclimatology}.
\newblock {\em Dendrochronologia}, 15:139--150.

\bibitem[Biondi, 2000]{Biondi2000}
Biondi, F. (2000).
\newblock {Are climate-tree growth relationships changing in north-central
  idaho, USA?}
\newblock {\em Arctic, Antarctic, and Alpine Research}, 32(2):111--116.

\bibitem[Breshears et~al., 2005]{Breshears2005}
Breshears, D.~D., Cobb, N.~S., Rich, P.~M., Price, K.~P., Allen, C.~D., Balice,
  R.~G., Romme, W.~H., Kastens, J.~H., Floyd, M.~L., and Belnap, J. (2005).
\newblock {Regional vegetation die-off in response to global-change-type
  drought}.
\newblock {\em Proceedings of the National Academy of Sciences of the United
  States of America}, 102(42):15144--15148.

\bibitem[Calder et~al., 2003]{Calder2003}
Calder, C., Lavine, M., M{\"u}ller, P., and Clark, J.~S. (2003).
\newblock Incorporating multiple sources of stochasticity into dynamic
  population models.
\newblock {\em Ecology}, 84(6):1395--1402.

\bibitem[Carlin et~al., 1992]{Carlin1992}
Carlin, B.~P., Polson, N.~G., and Stoffer, D.~S. (1992).
\newblock {A Monte Carlo approach to nonnormal and nonlinear state-space
  modeling}.
\newblock {\em Journal of the American Statistical Association},
  87(418):493--500.

\bibitem[Carrer and Urbinati, 2004]{Carrer2004}
Carrer, M. and Urbinati, C. (2004).
\newblock Age-dependent tree-ring growth responses to climate in larix decidual
  and pinus cembra.
\newblock {\em Ecology}, 85(3):730--740.

\bibitem[Carrer and Urbinati, 2006]{Carrer2006}
Carrer, M. and Urbinati, C. (2006).
\newblock Long-term change in the sensitivity of tree‐ring growth to climate
  forcing in larix decidua.
\newblock {\em New Phytologist}, 170(4):861--872.

\bibitem[Clark, 1989]{Clark1989}
Clark, J.~S. (1989).
\newblock {Effects of long-term water balances on fire regime, north-western
  Minnesota}.
\newblock {\em Journal of Ecology}, 77(4):989--1004.

\bibitem[Clark et~al., 2012]{Clark2012}
Clark, J.~S., Bell, D.~M., Kwit, M., Stine, A., Vierra, B., and Zhu, K. (2012).
\newblock {Individual-scale inference to anticipate climate-change
  vulnerability of biodiversity}.
\newblock {\em Philosophical Transactions of the Royal Society B: Biological
  Sciences}, 367(1586):236--246.

\bibitem[Clark and Bj{\o}rnstad, 2004]{Clark2004}
Clark, J.~S. and Bj{\o}rnstad, O.~N. (2004).
\newblock {Population time series: process variability, observation errors,
  missing values, lags, and hidden states}.
\newblock {\em Ecology}, 85(11):3140--3150.

\bibitem[Clark et~al., 2016]{Clark2016}
Clark, J.~S., Iverson, L., Woodall, C.~W., Allen, C.~D., Bell, D.~M., Bragg,
  D.~C., D'Amato, A.~W., Davis, F.~W., Hersh, M.~H., Ib{\'a}{\~n}ez, I., et~al.
  (2016).
\newblock The impacts of increasing drought on forest dynamics, structure, and
  biodiversity in the united states.
\newblock {\em Global Change Biology}, 22:2329--2352.

\bibitem[Clark et~al., 2007]{Clark2007}
Clark, J.~S., Wolosin, M., Dietze, M., Ib{\'a}{\~n}ez, I., LaDeau, S., Welsh,
  M., and Kloeppel, B. (2007).
\newblock {Tree growth inference and prediction from diameter censuses and ring
  widths}.
\newblock {\em Ecological Applications}, 17(7):1942--1953.

\bibitem[Cook, 1987]{Cook1987}
Cook, E.~R. (1987).
\newblock {The decomposition of tree-ring series for environmental studies}.
\newblock {\em Tree-Ring Bulletin}, 47:37--59.

\bibitem[Cook and Kairiukstis, 1990]{Cook1990}
Cook, E.~R. and Kairiukstis, L.~A., editors (1990).
\newblock {\em {Methods of Dendrochronology: Applications in the Environmental
  Sciences}}.
\newblock Kluwer Academic Publications, Hingham, MA.

\bibitem[Cook and Peters, 1981]{Cook1981}
Cook, E.~R. and Peters, K. (1981).
\newblock {The smoothing spline: a new approach to standardizing forest
  interior tree-ring width series for dendroclimatic studies}.
\newblock {\em Tree-ring bulletin}, 41:45--53.

\bibitem[Dale et~al., 2001]{Dale2001}
Dale, V.~H., Joyce, L.~A., Mcnulty, S., Neilson, R.~P., Ayres, M.~P.,
  Flannigan, M.~D., Hanson, P.~J., Irland, L.~C., Ariel, E., Peterson, C.~J.,
  Simberloff, D., Swanson, F.~J., Stocks, B.~J., Wotton, B.~M., Dale, V.~H.,
  Joyce, L.~A., Mcnulty, S., Ronald, P., Matthew, P., Simberloff, D., Swanson,
  F.~J., Stocks, B.~J., and Wotton, B.~M. (2001).
\newblock {Climate change and forest disturbances}.
\newblock {\em BioScience}, 51(9):723--734.

\bibitem[D'Amato et~al., 2013]{D'Amato2013}
D'Amato, A.~W., Bradford, J.~B., Fraver, S., and Palik, B.~J. (2013).
\newblock {Effects of thinning on drought vulnerability and climate response in
  north temperate forest ecosystems}.
\newblock {\em Ecological Applications}, 23(8):1735--1742.

\bibitem[Dyer, 2004]{Dyer2004}
Dyer, J.~M. (2004).
\newblock {A water budget approach to predicting tree species growth and
  abundance, utilizing paleoclimatology sources}.
\newblock {\em Climate Research}, 28(1):1--10.

\bibitem[Floyd et~al., 2009]{Floyd2009}
Floyd, M.~L., Clifford, M., Cobb, N.~S., Hanna, D., Delph, R., Ford, P., and
  Turner, D. (2009).
\newblock Relationship of stand characteristics to drought-induced mortality in
  three southwestern pi{\~{n}}on–juniper woodlands.
\newblock {\em Ecological Applications}, 19(5):1223--1230.

\bibitem[Foster et~al., 2014]{Foster2014}
Foster, J.~R., D'Amato, A.~W., and Bradford, J.~B. (2014).
\newblock {Looking for age-related growth decline in natural forests:
  Unexpected biomass patterns from tree rings and simulated mortality}.
\newblock {\em Oecologia}, 175(1):363--374.

\bibitem[Foster et~al., 2016]{Foster2016}
Foster, J.~R., Finley, A.~O., D'Amato, A.~W., Bradford, J.~B., and Banerjee, S.
  (2016).
\newblock Predicting tree biomass growth in the temperate–boreal ecotone: Is
  tree size, age, competition, or climate response most important?
\newblock {\em Global Change Biology}, 22(6):2138--2151.

\bibitem[Holmes, 1983]{Holmes1983}
Holmes (1983).
\newblock {Computer-assisted quality control in tree-ring dating and
  measurement}.
\newblock {\em Tree-ring bulletin}, 43(1):69.

\bibitem[Hooten and Hobbs, 2015]{Hooten2015}
Hooten, M.~B. and Hobbs, N.~T. (2015).
\newblock {A guide to Bayesian model selection for ecologists}.
\newblock {\em Ecological Monographs}, 85(1):3--28.

\bibitem[Hooten et~al., 2007]{Hooten2007}
Hooten, M.~B., Wikle, C.~K., Dorazio, R.~M., and Royle, J.~A. (2007).
\newblock {Hierarchical spatiotemporal matrix models for characterizing
  invasions}.
\newblock {\em Biometrics}, 63(2):558--567.

\bibitem[Innes and Cook, 1989]{Innes1989}
Innes, J.~L. and Cook, E.~R. (1989).
\newblock {Tree-ring analysis as an aid to evaluating the effects of pollution
  on tree growth}.
\newblock {\em Canadian Journal of Forest Research}, 19(9):1174--1189.

\bibitem[IPCC, 2013]{IPCC2013}
IPCC (2013).
\newblock {\em Climate Change 2013: The Physical Science Basis. Contribution of
  Working Group I to the Fifth Assessment Report of the Intergovernmental Panel
  on Climate Change}.
\newblock Cambridge University Press, Cambridge, United Kingdom and New York,
  NY, USA.

\bibitem[Kipfmueller et~al., 2010]{Kipfmueller2010}
Kipfmueller, K.~F., Elliott, G.~P., Larson, E.~R., and Salzer, M.~W. (2010).
\newblock An assessment of the dendroclimatic potential of three conifer
  species in northern minnesota.
\newblock {\em Tree-Ring Research}, 66(2):113--126.

\bibitem[Klos et~al., 2009]{Klos2009}
Klos, R.~J., Wang, G.~G., Bauerle, W.~L., and Rieck, J.~R. (2009).
\newblock {Drought impact on forest growth and mortality in the southeast USA:
  an analysis using Forest Health and Monitoring data}.
\newblock {\em Ecological Applications}, 19(3):699--708.

\bibitem[Laurent et~al., 2003]{Laurent2003}
Laurent, M., Antoine, N., and Jo{\"{e}}l, G. (2003).
\newblock {Effects of different thinning intensities on drought response in
  Norway spruce (Picea abies (L.) Karst.)}.
\newblock {\em Forest Ecology and Management}, 183(1):47--60.

\bibitem[Lutz et~al., 2010]{Lutz2010}
Lutz, J.~A., van Wagtendonk, J.~W., and Franklin, J.~F. (2010).
\newblock {Climatic water deficit, tree species ranges, and climate change in
  Yosemite National Park}.
\newblock {\em Journal of Biogeography}, 37(5):936--950.

\bibitem[Macalady and Bugmann, 2014]{Macalady2014}
Macalady, A.~K. and Bugmann, H. (2014).
\newblock {Growth-mortality relationships in pi{\~{n}}on pine (Pinus edulis)
  during severe droughts of the past century: shifting processes in space and
  time}.
\newblock {\em PloS one}, 9(5):e92770.

\bibitem[Mart{\'{\i}}nez-Vilalta et~al., 2012]{Martinez-Vilalta2012}
Mart{\'{\i}}nez-Vilalta, J., L{\'{o}}pez, B.~C., Loepfe, L., and Lloret, F.
  (2012).
\newblock {Stand-and tree-level determinants of the drought response of Scots
  pine radial growth}.
\newblock {\em Oecologia}, 168(3):877--888.

\bibitem[McDowell et~al., 2008]{McDowell2008}
McDowell, N., Pockman, W.~T., Allen, C.~D., Breshears, D.~D., Cobb, N., Kolb,
  T., Plaut, J., Sperry, J., West, A., and Williams, D.~G. (2008).
\newblock {Mechanisms of plant survival and mortality during drought: why do
  some plants survive while others succumb to drought?}
\newblock {\em New Phytologist}, 178(4):719--739.

\bibitem[McDowell et~al., 2006]{McDowell2006}
McDowell, N.~G., Adams, H.~D., Bailey, J.~D., Hess, M., and Kolb, T.~E. (2006).
\newblock {Homeostatic maintenance of ponderosa pine gas exchange in response
  to stand density changes}.
\newblock {\em Ecological Applications}, 16(3):1164--1182.

\bibitem[Millar et~al., 2007]{Millar2007}
Millar, C.~I., Stephenson, N.~L., and Stephens, S.~L. (2007).
\newblock {Climate change and forests of the future: managing in the face of
  uncertainty}.
\newblock {\em Ecological Applications}, 17(8):2145--2151.

\bibitem[Oliver and Larson, 1996]{Oliver1996}
Oliver, C.~D. and Larson, B.~C. (1996).
\newblock {\em Forest Stand Dynamics, Update Edition}.
\newblock John Wiley {\&} Sons, New York, NY.

\bibitem[Park and Casella, 2008]{Park2008}
Park, T. and Casella, G. (2008).
\newblock {The Bayesian Lasso}.
\newblock {\em Journal of the American Statistical Association},
  103(482):681--686.

\bibitem[Parslow et~al., 2013]{Parslow2013}
Parslow, J., Cressie, N., Campbell, E.~P., Jones, E., and Murray, L. (2013).
\newblock {Bayesian learning and predictability in a stochastic nonlinear
  dynamical model}.
\newblock {\em Ecological Applications}, 23(4):679--698.

\bibitem[{PRISM Climate Group}, 2013]{prism}
{PRISM Climate Group} (2013).
\newblock {Oregon State University, http://prism.oregonstate.edu, created 15
  Aug 2013}.

\bibitem[Puettmann, 2011]{Puettmann2011}
Puettmann, K.~J. (2011).
\newblock {Silvicultural challenges and options in the context of global
  change:“Simple” fixes and opportunities for new management approaches}.
\newblock {\em Journal of Forestry}, 109(6):321--331.

\bibitem[Reinikainen et~al., 2012]{Reinikainen2012}
Reinikainen, M., D'Amato, A.~W., and Fraver, S. (2012).
\newblock {Repeated insect outbreaks promote multi-cohort aspen mixedwood
  forests in northern Minnesota, USA}.
\newblock {\em Forest Ecology and Management}, 266:148--159.

\bibitem[Rozas et~al., 2009]{Rozas2009}
Rozas, V., DeSoto, L., and Olano, J.~M. (2009).
\newblock Sex-specific, age-dependent sensitivity of tree-ring growth to
  climate in the dioecious tree juniperus thurifera.
\newblock {\em New Phytologist}, 182(3):687--697.

\bibitem[Schliep et~al., 2014]{Schliep2014}
Schliep, E.~M., Dong, T.~Q., Gelfand, A.~E., and Li, F. (2014).
\newblock {Modeling individual tree growth by fusing diameter tape and
  increment core data}.
\newblock {\em Environmetrics}, 25(8):610--620.

\bibitem[Schofield et~al., 2016]{Schofield2015}
Schofield, M.~R., Barker, R.~J., Gelman, A., Cook, E.~R., and Briffa, K.~R.
  (2016).
\newblock A model-based approach to climate reconstruction using tree-ring
  data.
\newblock {\em Journal of the American Statistical Association},
  111(513):93--106.

\bibitem[Sohn et~al., 2016]{Sohn2016}
Sohn, J.~A., Hartig, F., Kohler, M., Huss, J., and Bauhus, J. (2016).
\newblock Heavy and frequent thinning promotes drought adaptation in pinus
  sylvestris forests.
\newblock {\em Ecological Applications}, page In press.

\bibitem[Stephenson, 1998]{Stephenson1998}
Stephenson, N.~L. (1998).
\newblock {Actual evapotranspiration and deficit: biologically meaningful
  correlates of vegetation distribution across spatial scales}.
\newblock {\em Journal of Biogeography}, 25:855--870.

\bibitem[Van~Deusen, 1987]{VanDeusen1987}
Van~Deusen, P.~C. (1987).
\newblock {Testing for stand dynamics effects on red spruce growth trends}.
\newblock {\em Canadian Journal of Forest Research}, 17(12):1487--1495.

\bibitem[{Van Deusen}, 1989]{VanDeusen1989}
{Van Deusen}, P.~C. (1989).
\newblock {A model-based approach to tree ring analysis}.
\newblock {\em Biometrics}, 45(3):763--779.

\bibitem[Visser, 1986]{Visser1986}
Visser, H. (1986).
\newblock {Analysis of tree ring data using the Kalman filter technique}.
\newblock {\em IAWA Bulletin n.s.}, 7(4):289--297.

\bibitem[Visser and Molenaar, 1988]{Visser1988}
Visser, H. and Molenaar, J. (1988).
\newblock {Kalman filter analysis in dendroclimatology}.
\newblock {\em Biometrics}, 44:929--940.

\bibitem[West and Harrison, 1997]{West1997}
West, M. and Harrison, J. (1997).
\newblock {\em {Bayesian Forecasting and Dynamic Models}}.
\newblock Springer Series in Statistics. Springer New York.

\bibitem[Wikle, 2003]{Wikle2003}
Wikle, C.~K. (2003).
\newblock {Hierarchical Bayesian models for predicting the spread of ecological
  processes}.
\newblock {\em Ecology}, 84(6):1382--1394.

\bibitem[Williams et~al., 2010]{Williams2010}
Williams, A.~P., Allen, C.~D., Millar, C.~I., Swetnam, T.~W., Michaelsen, J.,
  Still, C.~J., and Leavitt, S.~W. (2010).
\newblock {Forest responses to increasing aridity and warmth in the
  southwestern United States}.
\newblock {\em Proceedings of the National Academy of Sciences},
  107(50):21289--21294.

\bibitem[Wood, 2006]{Wood2006}
Wood, S. (2006).
\newblock {\em {Generalized Additive Models: an Introduction with R}}.
\newblock CRC press, Boca Raton, FL.

\bibitem[Yamaguchi, 1991]{Yamaguchi1991}
Yamaguchi, D.~K. (1991).
\newblock {A simple method for cross-dating increment cores from living trees}.
\newblock {\em Canadian Journal of Forest Research}, 21(3):414--416.

\end{thebibliography}
\bibliographystyle{apalike}

\end{document}